\documentclass[a4paper,12pt]{article}
\usepackage{graphicx}
\usepackage{bm}
\usepackage{authblk}
\usepackage{xcolor}
\usepackage{subcaption}
\usepackage{amsmath}
\usepackage{amsfonts}
\usepackage[normalem]{ulem}
\usepackage{setspace}
\usepackage{tikz}
\usepackage{cite}
\usepackage{ulem}
\usepackage{url}
\usepackage[font=small]{caption}
\usepackage{cite}   
\usepackage{array}
\usepackage{ragged2e}
\usepackage[utf8]{inputenc}
\usepackage[T1]{fontenc}

\justifying
\newcolumntype{P}[1]{>{\centering\arraybackslash}p{#1}}

\begin{document}


\title{Two octave supercontinuum generation in a non-silica graded-index multimode fiber}
\date{}

\author[1]{\small Zahra Eslami}
\author[1]{\small Lauri Salmela}
\author[2,3]{\small Adam Filipkowski}
\author[2]{\small Dariusz Pysz}
\author[3]{\small Mariusz Klimczak}
\author[2,3]{\small Ryszard Buczynski}
\author[4]{\small John M. Dudley}
\author[1,*]{\small Go{\"e}ry Genty}

\affil[1]{\footnotesize Photonics Laboratory, Physics Unit, Tampere University, 33014 Tampere, Finland}
\affil[2]{\footnotesize Łukasiewicz Research Network – Institute of Microelectronics and Photonics, Al. Lotników 32/46, 02-668 Warsaw, Poland}
\affil[3]{\footnotesize University of Warsaw, Faculty of Physics, Pasteura 5, 02-093 Warsaw, Poland.}
\affil[4]{\footnotesize Institut FEMTO-ST, Universit\'{e} Bourgogne Franche-Comt\'{e} CNRS UMR 6174, 25000 Besan\c{c}on, France}
\affil[*]{goery.genty@tuni.fi}

\maketitle

\newcommand{\jdnote}[1] {{\textcolor{black}{#1}}} 





 


\begin{abstract}

The generation of a two-octave supercontinuum from the visible to mid-infrared (700--2800~nm) in a non-silica graded-index multimode fiber is reported. The fiber design is based on a nanostructured core comprised of two types of drawn lead-bismuth-gallate glass rods with different refractive indices. This structure yields an effective parabolic index profile, an extended transmission window, and ten times increased nonlinearity when compared to silica fibers. Using femtosecond pulse pumping at wavelengths in both normal and anomalous dispersion regimes, a detailed study is carried out into the supercontinuum generating mechanisms and instabilities seeded by periodic self imaging. Significantly, suitable injection conditions in the high power regime are found to result in the output beam profile showing clear signatures of beam self-cleaning from nonlinear mode mixing.  Experimental observations are interpreted using spatio-temporal 3+1D numerical simulations of the generalized nonlinear Schr{\"o}dinger equation, and simulated spectra are in excellent agreement with experiment over the full two-octave spectral bandwidth. These results demonstrate a new pathway towards the generation of bright, ultrabroadband light sources in the mid-infrared.

\end{abstract}


\section{Introduction}
Understanding the physics of complex nonlinear optical systems has been the focus of intense research in recent years and, in this context, the generation of broadband supercontinuum in graded-index multimode fibers (GRIN MMFs) has attracted particular attention \cite{renninger2013optical,wright2015spatiotemporal,ahsan2018graded,wright2015controllable,krupa2016observation,arabi2018geometric,agrawal2019invite,krupa2019multimode}.  \jdnote{In addition to multimode fibers providing additional degrees of freedom to optimize  supercontinuum for specific applications, the spatio-temporal propagation in such fibers reveals a rich landscape of nonlinear dynamics, with close links to universal phenomena such as wave turbulence \cite{garnier2019wave,baudin2020classical,malkin2018transition,podivilov2019hydrodynamic}.}

In contrast to single-mode fibers where the spatial intensity distribution remains constant with propagation, the parabolic index profile of GRIN fibers leads to a periodic self-imaging phenomenon enabling \jdnote{spatio-temporal coupling and mode mixing associated with complex and unique nonlinear dynamics.} \jdnote{Amongst spatio-temporal effects that have been observed include} the generation of multimode solitons \cite{yu1995spatio,raghavan2000spatiotemporal,renninger2013optical,wright2015spatiotemporal,ahsan2018graded}, the development of geometric parametric instabilities \cite{wright2015controllable,krupa2016observation,arabi2018geometric,agrawal2019invite}, and the formation of GRIN lenses \cite{filipkowski2016world,hudelist2009design}. Spatio-temporal dynamics in GRIN silica fibers have been exploited to manipulate the transverse beam profile \cite{wright2015controllable, deliancourt2019wavefront} and under particular injection conditions, the output beam was observed to exhibit a quasi single-mode profile as the result of nonlinear self-cleaning dynamics \cite{liu2016kerr,krupa2017spatial,laegsgaard2018spatial,hansson2020nonlinear}. Numerical studies have shown that such self-cleaning is a particular feature of pulse propagation in GRIN MMFs \cite{mafi2012pulse, conforti2017fast, sidelnikov2019random} associated with strong nonlinear coupling leading to preferential energy transfer to the low-order modes \cite{wright2016self,krupa2017spatial}. Beam self-cleaning has been reported under various pumping configurations using nanosecond \cite{wright2016self}, picosecond \cite{deliancourt2019kerr, krupa2017spatial} and femtosecond \cite{liu2016kerr} pulses, both in the normal \cite{liu2016kerr,lopez2016visible,krupa2016observation,krupa2016spatiotemporal} and anomalous dispersion regime \cite{eftekhar2017versatile}. \jdnote {In addition to being of fundamental importance through its links to universal nonlinear physics, dynamical self-cleaning is also of great practical interest in the development of high power supercontinuum (SC) sources.}

\jdnote{To date, however, all studies and demonstration of SC generation in graded-index fibers have been restricted to silica fibers, and bandwidths limited only to the visible and near-infrared spectral regions \cite{deliancourt2019kerr, leventoux2018kerr, liu2016kerr, eftekhar2017versatile, lopez2016visible}. Yet because of the ability of GRIN MMFs to provide power scaling with a near-Gaussian spatial intensity distribution, there is major interest in extending the GRIN MMF platform into the mid-infrared regime where high spatial beam quality and high power  \cite{tu2013coherent} are required in applications 
 including e.g. molecular fingerprinting \cite{petersen2014mid}, microscopy\cite{dupont2012ir,kilgus2018diffraction}, medical diagnostics \cite{seddon2017biomedical, seddon2018prospective}, gas monitoring \cite{werle2002near, grassani2019mid}, spectroscopy \cite{kumar2012stand, amiot2017cavity}, optical coherence tomography \cite{petersen2018mid} and LIDAR \cite{saleh2019short}.  }

Here, we fill this gap and report the generation of a two-octave supercontinuum expanding from 700~nm to 2800~nm in a non-silica graded-index multimode fiber. The fiber is designed using two types of lead-bismuth-gallate (PBG) glass rods with different refractive indices drawn to yield a nanostructured core \cite{sobon2014infrared,stepien2013development, cimek2017experimental}. The result is a multimode fiber with an effective parabolic refractive index profile, enhanced nonlinearity and transmission window up to 2800 nm. Injecting femtosecond pulses in the fiber, we observe the generation of a two-octave supercontinuum from 700 to 2800 nm. We conduct a systematic investigation of the SC generating mechanism as a function of pump wavelength, with self-phase modulation (SPM) and geometric parametric instabilities (GPI) seeding the SC generation process in the normal dispersion regime, while in the anomalous regime soliton dynamics and parametric dispersive waves excitation are found to dominate. The relative intensity noise \jdnote{has also been} characterized in several wavelength bands across the SC spectrum and, under particular controlled injection conditions, we see clear signatures of self-cleaning dynamics with a near single-mode spatial intensity distribution at the fiber output. \jdnote{In order to confirm and interpret our experiments,} we perform spatio-temporal 3+1D numerical simulations of the generalized nonlinear Schr{\"o}dinger equation. Remarkably, the simulations reproduce the spatial intensity distributions measured at the fiber output and confirm the self-cleaning dynamics, with spectra in excellent agreement over the full SC bandwidth. \jdnote{These} results not only open up novel perspectives for the study of nonlinear spatio-temporal instabilities in non-silica graded-index fiber platforms but also provide an avenue for power-scaling of supercontinuum sources in the mid-infrared.

\section{Fiber \jdnote{design} characteristics}

\begin{figure}[h]
  \includegraphics[width=\linewidth]{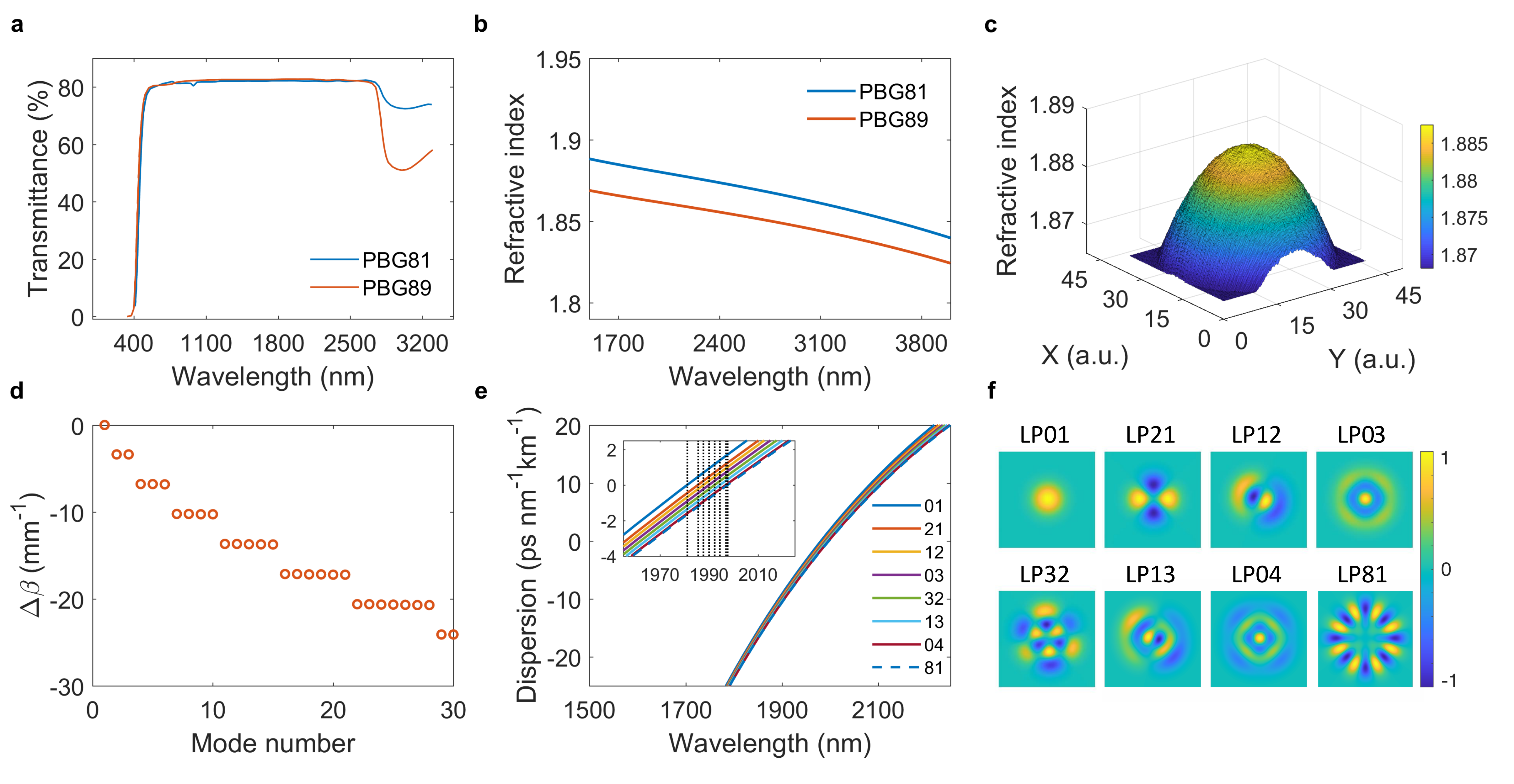}
  \caption{Lead-bismuth-gallate (PBG) graded-index multimode optical ﬁber characteristics. a) Transmission of the two PBG81 (blue) and PBG89 (red) constituent glasses. b) Refractive index of PBG81 (blue) and PBG89 (red) calculated using the Sellmeier model. c) Simulated 3D average refractive index distribution of the fiber structure. d) Propagation constant of the first 30 modes of the fiber with 80~$\mu m$ core size and refractive-index profile as in c. e) Dispersion profile of the PBG fiber fundamental and selected higher-order modes. The inset in Fig.~\ref{fig:refractive-index}e shows a magnified view of the dispersion profile near the zero-dispersion region where the vertical dashed lines mark the position of the ZDW. f) Spatial amplitude distribution corresponding to the dispersion profile of the modes shown in e.}
  \label{fig:refractive-index}
\end{figure}

The fiber preform was designed to have a parabolic index profile using two types of in-house developed lead-bismuth-gallate (PBG) glasses (see Methods for details on the fabrication process). Figure~\ref{fig:refractive-index} shows the characteristics of the fabricated GRIN PBG fiber with $\rm R=40~\rm{\mu m}$ core radius. The relative index difference $\Delta = (n_{co}-n_{cl})/n_{co}$=0.0101 corresponds to a numerical aperture $\rm NA=0.26$ with $n_{co}$ and $n_{cl}$ the refractive index at the core center and in the cladding, respectively. Both glasses have a transmission window extending from 400 nm to about 2800 nm limited by OH absorption (Fig.~\ref{fig:refractive-index}a) and a refractive index value close to 1.85 (Fig.~\ref{fig:refractive-index}b). The simulated refractive index distribution of the designed fiber is shown in Fig.~\ref{fig:refractive-index}(c) and exhibits a parabolic variation from the cladding to the core center.

Figure~\ref{fig:refractive-index}(d) shows the simulated propagation constant $\Delta\beta$ of the first 30 modes of the fiber relative to that of the fundamental mode. In contrast to step-index fibers, one can see how the parabolic refractive index profile yields discrete clusters of modes with equally spaced propagation constant and with a cluster population that grows with $\Delta\beta$. This particular feature results in periodic multimodal interference and self-imaging, where the spatial field focuses and defocuses periodically with a period $z_p=\pi R/\sqrt{2\Delta}$ \cite{agrawal2019invite}. 

In the experiments reported below, we operate in the regime where the injected peak power is below the critical power for catastrophic self-focusing such that the self-imaging condition is essentially independent of the nonlinear Kerr contribution (see Supplementary Information). Modes within a particular cluster experience identical propagation constant resulting in minimum modal dispersion and walk-off \cite{wright2017multimode}. The propagation constants difference between modes of distinct clusters is large, leading to strong intra-cluster modal interaction but limited inter-cluster mode coupling, such that the fundamental mode is the most stable propagating mode. We also numerically simulated the group velocity dispersion (GVD) of the fundamental and selected higher-order modes vs. wavelengths as shown in Fig.~\ref{fig:refractive-index}(e) with the corresponding spatial amplitude of the modes illustrated in Fig.~\ref{fig:refractive-index}(f). Unlike in step-index multimode fibers \cite{eslami2019high}, the GVD characteristics of the different modes do not differ significantly with a near-constant zero dispersion wavelength (ZDW) at $\sim$ 1980~nm for all modes. 


\section{Experiments and Results}
\begin{figure}[h]
  \includegraphics[width=\linewidth]{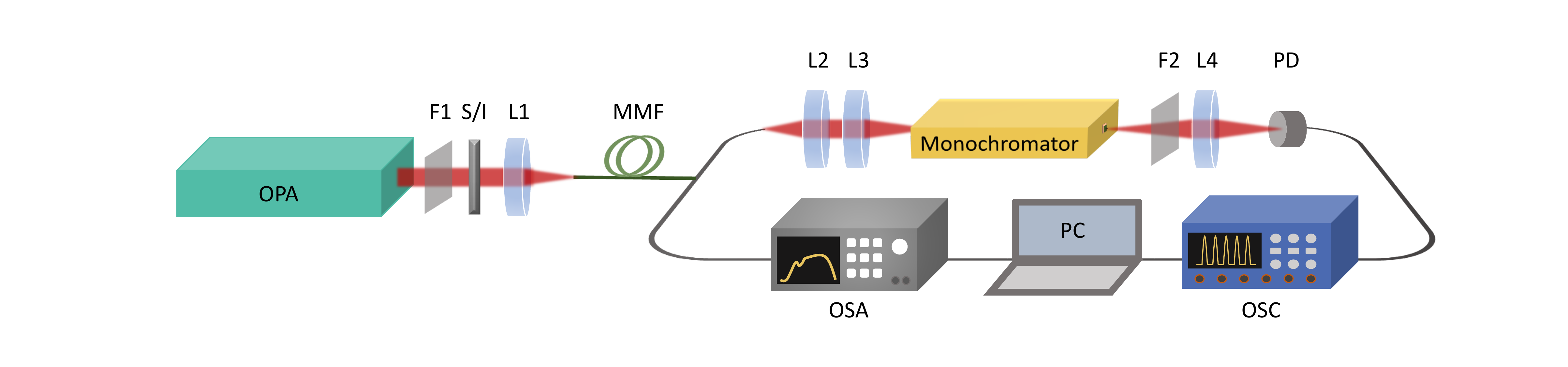}
  \caption{Schematic illustration of the experimental setup for SC generation and intensity noise measurement. OPA, optical parametric amplifier; L1-L4, plano-convex lenses; F1-F2, long-pass filters; S/I,  dichroic filter to select signal/idler; PD, photodetector; MMF, multimode fiber; OSA, optical spectrum analyzer; OSC, oscilloscope. The inset shows the measured intensity fluctuations of the OPA at 1700 nm.}
  \label{fig:setup}
\end{figure}

A schematic  illustration of our experimental setup is shown in Fig.~\ref{fig:setup} (see also Methods for additional details). A tunable optical parametric amplifier (OPA) producing 350 fs pulses at a repetition rate of 500 kHz is used as the pump source. The PBG fiber is 20 cm long. Two different optical spectrum analyzers are used to measure the SC spectrum in different wavelength ranges and a monochromator was used to filter out selected wavelength bands and characterize the pulse-to-pulse fluctuations (see Methods for details). The spatial intensity at the fiber output was characterized in the far-field using a beam profiling camera.

\begin{figure}
  \includegraphics[width=\linewidth]{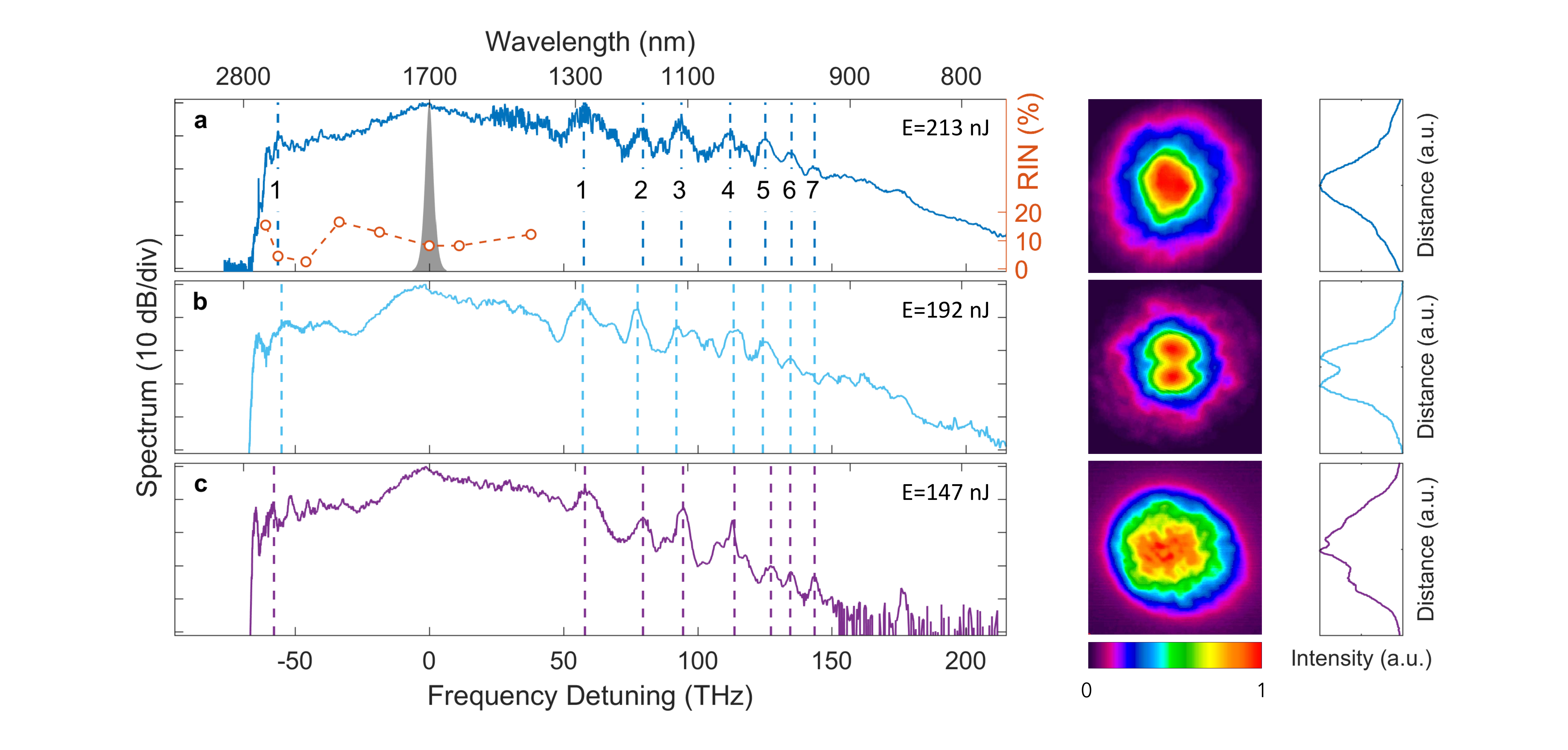}
  \caption{Experimentally measured spectra and corresponding far-field spatial intensity distributions recorded at the output of the GRIN PBG fiber for three different initial launching conditions in a, b and c. The vertical dashed lines correspond to the GPI sidebands frequencies with the order as indicated. The red circles shows the relative intensity noise (RIN) measured for 4000 consecutive pulses in different wavelength bands with 6 nm bandwidth (see Methods).}
  \label{fig:SC_excitation}
\end{figure}

We first characterized the output spectrum and spatial intensity distribution at the fiber output for input pulses at 1700~nm in the normal dispersion regime of the fiber. These results are shown in Fig.~\ref{fig:SC_excitation} for three different injection conditions corresponding to increasing tilt between the input beam and the fiber input facet. In all cases, the SC spans from 800~nm to 2800~nm (-40 dB bandwidth) with apparent discrete spectral components on the short wavelength side. These arise from the nonlinear refractive index grating induced by the Kerr effect and periodic self-imaging along propagation \cite{longhi2003modulational, krupa2016observation,mangini2021experimental} that exponentially amplifies small perturbations in frequency bands determined by a specific phasematching condition (see Methods). In contrast to conventional modulation instability in single mode fiber which is a pure temporal effect restricted to the anomalous dispersion regime, geometric parametric instability (GPI) is a spatio-temporal phenomenon that can occur regardless of the dispersion sign \cite{agrawal2019invite}. The bandwidth and amplitude of the GPI sidebands decrease with the order, in agreement with the theory \cite{longhi2003modulational}. In principle, GPI leads to the generation of multiple sideband pairs which are widely separated and symmetrically located around the pump. Here, however, because the long wavelength components fall outside the transmission window of the fiber, we only observe the short wavelength sidebands except for the first order where we see both spectral components. Note that this was also the case in previous studies in silica GRIN fibers \cite{lopez2016visible,krupa2016observation}. 

In order to confirm the origin of the discrete spectral components, we computed the theoretical location of the GPI sidebands for our fiber, and they are shown as dotted lines in Fig.~\ref{fig:SC_excitation} with the sideband order as indicated. For completeness, the experimental GPI high frequency components and corresponding analytical values up to the seventh order are listed in Table~\ref{table:1} where we see excellent agreement for all generated GPI orders.

\begin{table}[h!]
\begin{center}
\begin{tabular}{ l P{0.8cm} P{0.8cm} P{0.8cm} P{0.8cm} P{0.8cm} P{0.8cm} P{0.8cm} }
\hline
 \hspace{-0.22cm}\begin{tabular}{l}GPI order\\ \hline GPI Sidebands [THz]\end{tabular} & 1 & 2  & 3 & 4 & 5  & 6 & 7\\
 \hline
 Theory 
  & 54  &  77 &  95  & 110 & 123  & 134  & 145\\
 Experiment (a)  & 57  &  79  & 94  & 112 &  125 &  135 &  144\\
 Experiment (b) & 57  &  78  & 92  & 113 &  124 &  135 &  144\\
 Experiment (c)  & 58  &  80  & 95  & 114 &  127 &  135 &  144\\ \hline
\end{tabular}
\end{center}
\caption{Comparison of theoretically calculated (see Methods) and experimentally measured GPI sideband frequencies.}
\label{table:1}
\end{table}

Similarly to conventional modulation instability, an important feature of the geometric parametric instability observed here is that it is seeded by noise outside the pump spectral bandwidth. This is expected to lead to significant shot-to-shot fluctuations yielding a relatively smooth SC spectrum without fine structure when measured with an integrating optical spectrum analyzer \cite{dudley2006supercontinuum}. Intensity fluctuations were characterized in selected wavelength bands indicated by the circles in Fig.~\ref{fig:SC_excitation}a (see Methods). One can see that the fluctuations are minimal near the pump at 1700 nm and increase significantly away from the pump residue confirming the influence of noise amplification in the SC development. 

\begin{figure}[t]
  \includegraphics[width=\linewidth]{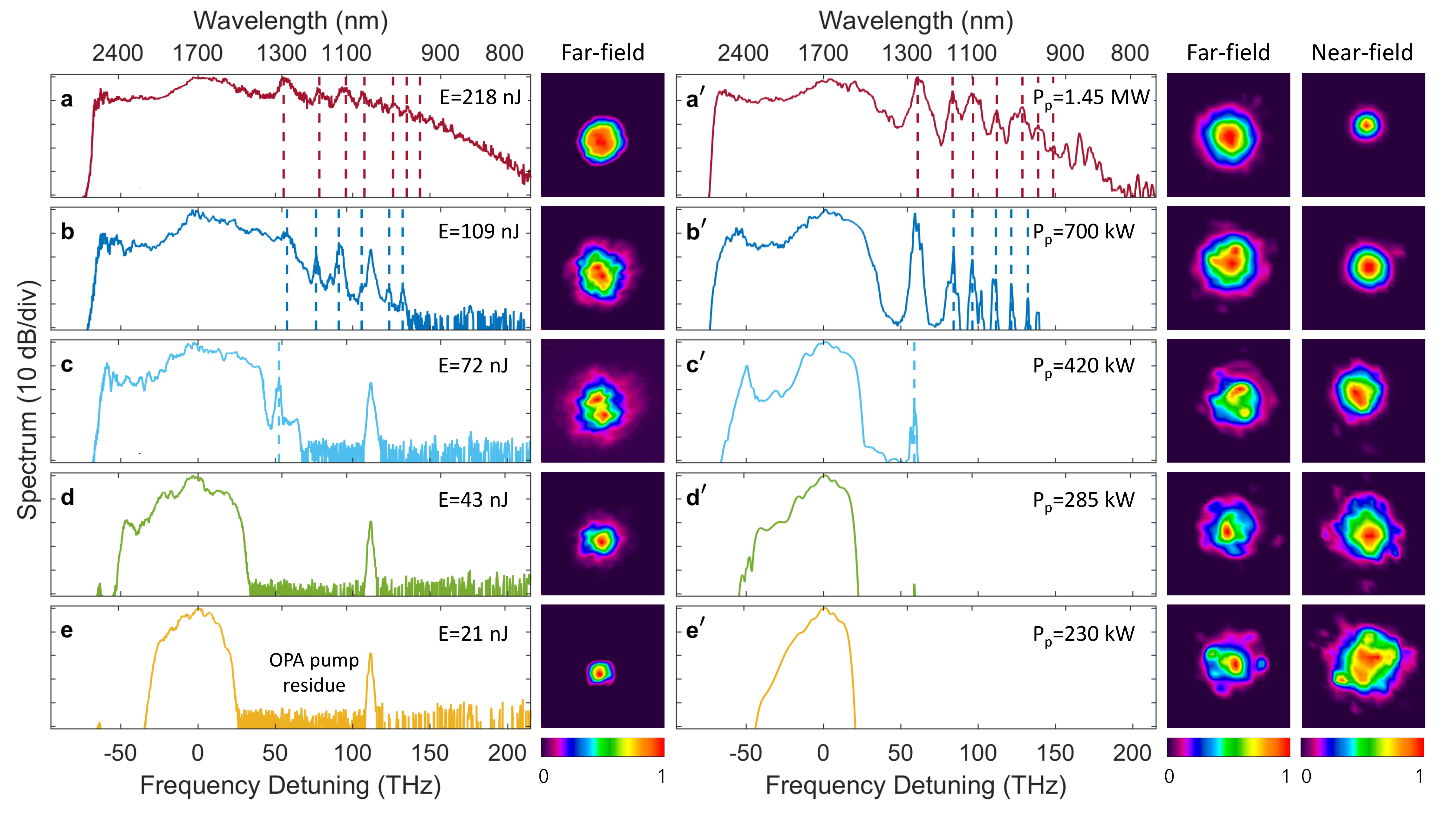}
  \caption{Supercontinuum spectra and corresponding far-field spatial intensity distribution as a function of injected energy for a pump wavelength of 1700 nm. a) E=218 nJ b) E=109 nJ c) E=72 nJ d) E=43 nJ e) E=21 nJ. $a^\prime$ to $e^\prime$ corresponding 3+1D numerically simulated SC spectra, far field and near field spatial intensity distributions. The vertical dashed lines marks the GPI sidebands frequencies. The spectral peak at a relative detuning of 110 THz (1060 nm) is a residue of the OPA pump laser.}
  \label{Power_dependent_SC}
\end{figure}

The spatial intensity profile at the fiber output corresponding to the spectra in Fig.~\ref{fig:SC_excitation} and obtained for different initial excitation conditions shows signatures of self-cleaning dynamics similar to previous observations in silica fibers \cite{wright2016self,krupa2017spatial,liu2016kerr,lopez2016visible,hansson2020nonlinear,deliancourt2019kerr}. Specifically, the fiber theoretically supports $\sim$750 transverse modes at 1700~nm, however when light is injected at normal incidence (Fig.~\ref{fig:SC_excitation}a) the measured spatial intensity distribution at the fiber output displays a quasi-Gaussian profile close to the fundamental $LP_{01}$ mode. When light was injected with a small angle with respect to the fiber axis in order to reduce the fraction of energy coupled to the fundamental mode, the intensity profile at the fiber output showed two distinct side-lobes ( Fig.~\ref{fig:SC_excitation}b) characteristic of the $LP_{11}$ mode. Increasing the input angle further to excite a larger fraction of modes at the fiber input (Fig.~\ref{fig:SC_excitation}c) yielded an output intensity profile with fine speckle-like structure indicative of multiple modes contribution. These observations are consistent with nonlinear mode-mixing dynamics mediated by the nonlinear refractive index grating induced along the propagation direction \cite{krupa2017spatial, hansson2020nonlinear}. Although the number of modes excited by the injection condition remains essentially constant with propagation \cite{podivilov2019hydrodynamic}, higher-order modes transfer energy via strong nonlinear coupling towards preferential modes as the result of optical wave turbulence \cite{malkin2018transition,podivilov2019hydrodynamic}. This energy flow is similar to wave condensation observed in hydrodynamic and depends on the initial spatial overlap between the input beam and fiber modes \cite{eftekhar2017versatile,malkin2018transition,garnier2019wave,podivilov2019hydrodynamic}. In the case of normal incidence where the input energy is distributed among the lowest-order modes and concentrated around the fiber longitudinal axis, energy preferentially flows from unstable higher-order modes to the fundamental mode \cite{hansson2020nonlinear}, while for a small input tilt the off-axis periodic local intensity oscillation generates an off-axis refractive-index grating overlapping with the $LP_{11}$ mode \cite{deliancourt2019kerr}. For a larger tilt, the fraction of energy coupled to higher-order modes at the fiber input is too large for self-cleaning dynamics to fully develop \cite{podivilov2019hydrodynamic}. Interestingly, and although the coupling efficiency decreases for larger input tilt, one can see that the output SC spectrum is nearly independent of the launching conditions. This can be seen from the spectral bandwidth as well as from the central frequencies of the GPI sidebands which remain unchanged in all three cases illustrated. This can be attributed to the fact that the dispersion profile of all modes is nearly identical, leading to dynamics which are essentially independent of the number of excited modes.


In order to study the influence of the spatio-temporal dynamics on the supercontinuum spectrum and spatial intensity profile at the fiber output, we performed additional measurements gradually increasing the input peak power for normal incidence launching condition similar to that in Fig.~\ref{fig:SC_excitation}a. The corresponding measured SC spectra and far-field intensity profiles are plotted in Fig.~\ref{Power_dependent_SC}. At the lowest injected power value (for which our camera can image the spatial intensity distribution), the nonlinear dynamics are dominated by SPM with near-symmetrical spectral broadening. The corresponding far-field spatial intensity profile is smooth with a narrow diameter, consistent with previous observations in the femtosecond regime \cite{liu2016kerr}. As the injected peak power is increased, discrete short-wavelength GPI sidebands develop from noise and the SPM-broadened spectrum extends into the anomalous dispersion regime where soliton dynamics \jdnote{develop.} Concomitantly in the far-field, we see an increase in diameter of the beam profile with an apparent contribution from low-order modes. For an input peak power in excess of 1 MW, the beam profile shows a dominant contribution from the $LP_{01}$ mode. 

To corroborate our experimental observations, we performed numerical simulations using the 3+1D generalized nonlinear Schr{\"o}dinger equation (see Methods for details) and the results are shown in the right sub-panels of Fig.~\ref{Power_dependent_SC}. The input peak power values in the simulations are adjusted to yield energies at the fiber output similar to those measured in our experiments. We also emphasize that, for the fiber studied here, Raman-induced dynamics are limited because of the significantly reduced Raman contribution arising both from the lower amplitude of the Raman gain in silicate glasses and the short fiber length (see Supplementary Information). The experimentally measured far-field intensity profiles corresponds to the Fourier transform of the near field and for completeness our simulations shows both near and far-field distributions. Remarkably, one can see overall excellent agreement between both the experimentally measured and simulated spectra and intensity profiles for all peak power values, with SPM-dominated broadening and the emergence of GPI-induced discrete spectral components widely separated from the pump at lower power. For sufficient increase in injected power, a fraction of the energy is transferred to the anomalous dispersion region with the formation of solitons. At the highest power values, cross-phase modulation interaction between the SPM spectral component, GPI sidebands and ejected solitons leads to a quasi-continuous SC spectrum. This scenario is confirmed in the spectrogram animation shown in Supplementary Movie 1 based on a simplified 1+1D model \cite{conforti2017fast} which reproduce the essential features of the nonlinear propagation (see Supplementary Information). We also see how nonlinear mode mixing dynamics leads to smoother near-field spatial intensity distribution with decreased beam size. 

\begin{figure}[!h]
  \includegraphics[width=\linewidth]{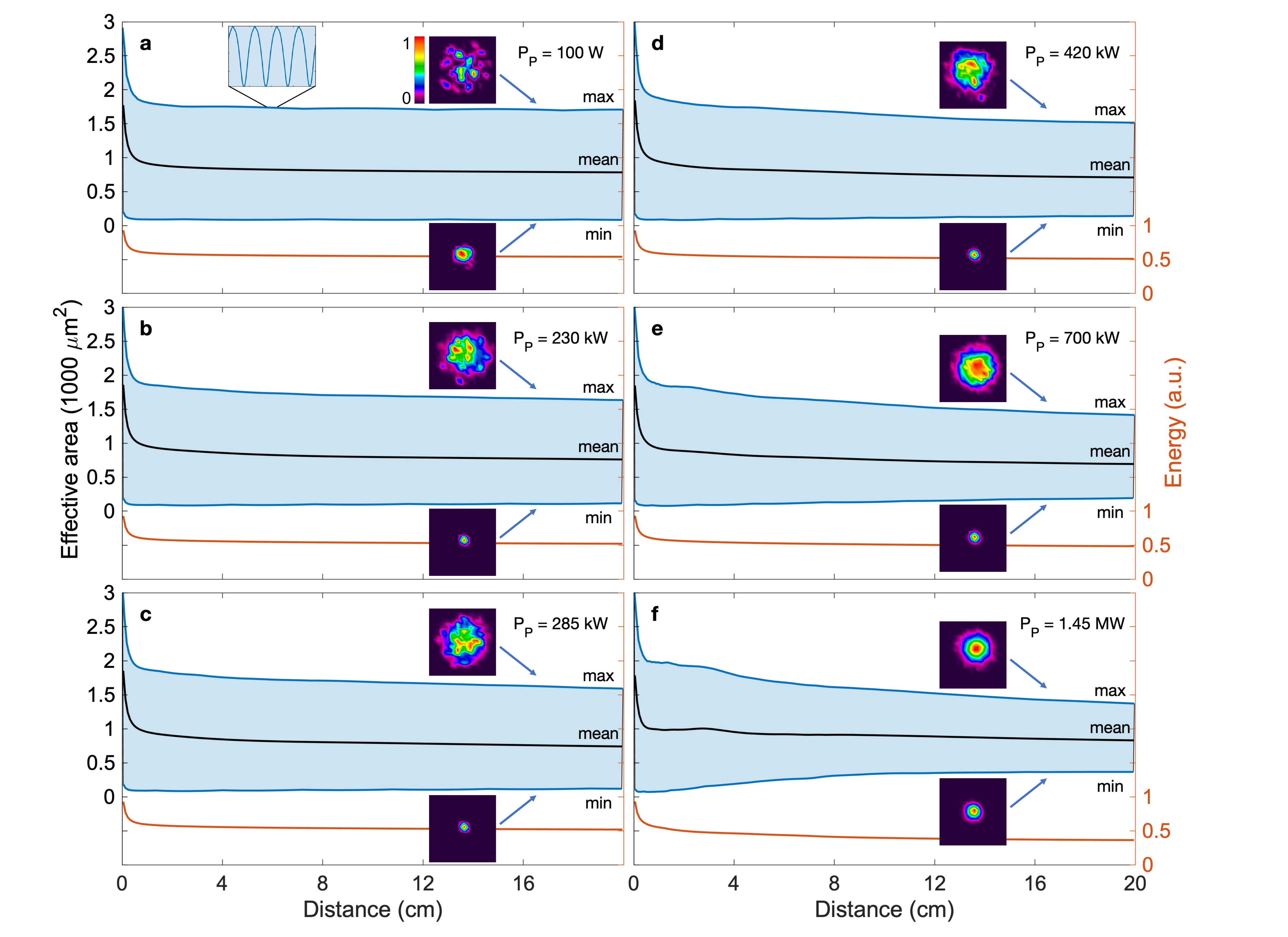}
  \caption{Simulated effective area variations from the linear to nonlinear propagation regime with injected peak power as indicated. The blue solid lines (min and max) show the envelope of the effective area fluctuations due to self-imaging. The black solid line corresponds to the mean effective area calculated over one self-imaging period. Inset in (a) shows the effective area fluctuations inside the envelope over 4 self-imaging periods. For each injected peak power, examples of near-field spatial intensity distributions corresponding to the minimum and maximum beam size are also illustrated at a propagation distance of 17~cm. The right axis (orange line) on each sub-panel shows the change in normalized energy as the result of leakage and infrared absorption along propagation with cumulative losses of 2.7, 2.8, 2.9, 2.9, 3.2 and 4.4~dB for a-f, respectively.}
  \label{fig:aeff}
\end{figure}

The role of self-cleaning dynamics at higher peak power values are further highlighted in Fig. \ref{fig:aeff} where we plot the simulated evolution of the beam effective area along the fiber for the different spectra shown in Fig.~\ref{Power_dependent_SC}. For comparison we also include the evolution in the linear regime in the absence of spectral broadening dynamics. In all cases, the effective area oscillates periodically as the result of the self-imaging dynamics (see inset) and it is this periodic perturbation that leads to the phasematched generation of parametric sidebands \cite{longhi2003modulational,agrawal2019invite}. We also see that the spatial intensity distribution varies significantly with propagation with clear contributions from higher-order modes in the defocused regions for lower injected peak power (Fig. \ref{fig:aeff}a-d). As the injected peak power is significantly increased, one can see how the contribution from the higher-order modes decreases dramatically as the result of nonlinear mode mixing with self-cleaning already occuring in the first few centimeters of the fiber (Fig. \ref{fig:aeff}e-f).  

\begin{figure}[hbt]
  \includegraphics[width=\linewidth]{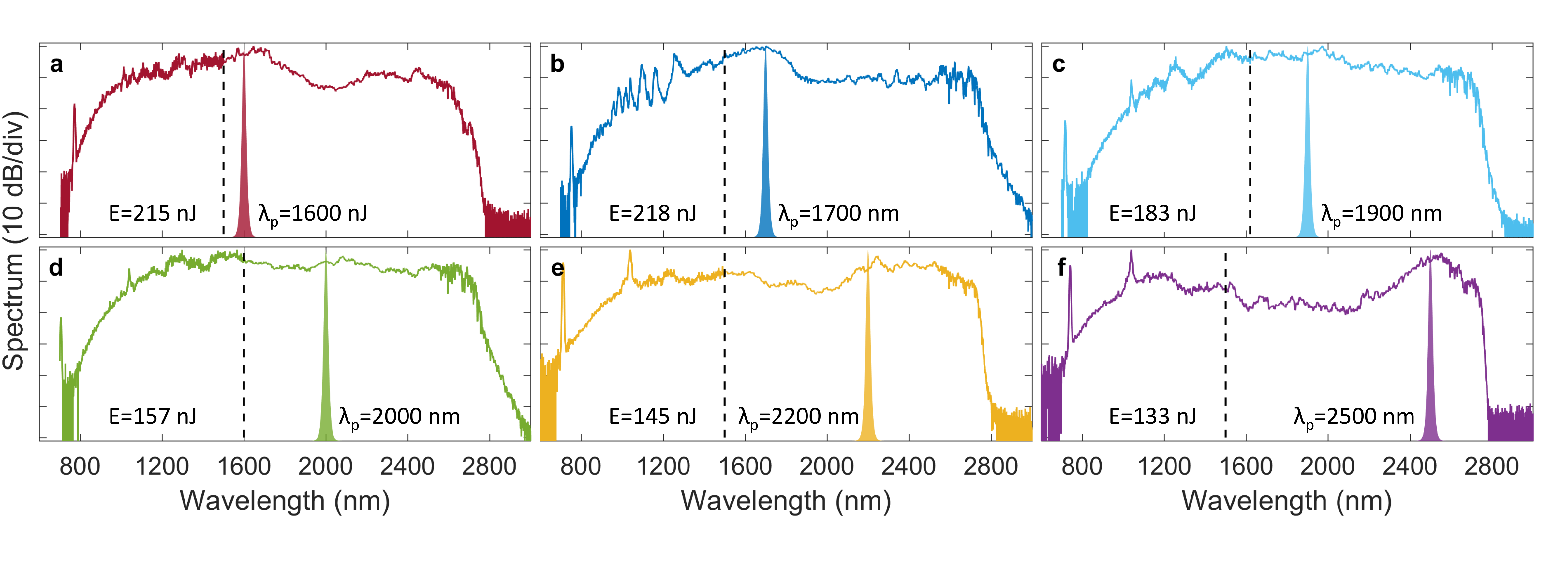}
  \caption{Experimentally measured SC spectra obtained in the 20 cm PBG fiber with pump wavelength at a) 1600~nm b) 1700~nm c) 1900~nm d) 2000~nm e) 2200~nm and f) 2500~nm with corresponding injected pulse energy $E$ as indicated in each plot. The OPA spectrum is also shown for each pump wavelength and the the dashed line marks the spectral regions measured using the different OSAs. The spectral peaks at wavelengths of 800 nm and 1060 nm are residues of the OPA and OPA pump laser, respectively.}
  \label{fig:spectra}
\end{figure}

We next investigated the influence of the pump wavelength on the SC spectrum. The results are illustrated in Fig.~\ref{fig:spectra} where we plot the normalized spectra measured at the fiber output corresponding to normal incidence injection conditions that maximized the SC bandwidth. The SC bandwidth is essentially independent of the pump wavelength, with the resulting SC spectrum spanning from 700-800 nm to 2800 nm in all cases, limited in the long wavelengths side by the fiber intrinsic high attenuation. Interestingly, the GPI sidebands only appear visible in the spectrum when the pump is located in the normal dispersion regime (Fig.~\ref{fig:spectra}a-c) but they are not present when the pump is tuned to the anomalous dispersion region beyond 2000 nm (Fig.~\ref{fig:spectra} d-f). 

This difference can be explained in the light of the different SC generating dynamics at play in the two regimes. Specifically, in the anomalous dispersion regime, the initial stage of propagation is dominated by higher-order soliton compression and fission accompanied by the generation on the short wavelength side of multiple dispersive waves phasematched by the self-imaging dynamics (see Supplementary Information and the spectrogram animation presented in the Supplementary Movie 2). At lower power values the dispersive wave components are clearly visible in the spectrum (see Supplementary Information) but when the injected power is significantly increased, interaction between the ejected solitons and dispersive waves leads to broadening of the dispersive wave components and their merging. 


\section{Conclusions}
Despite significant progress in mid-infrared SC generation, there are still important challenges to be overcome in order to obtain SC spectra with characteristics similar to those achieved in the visible/near-infrared. Silica-based platforms exhibit large absorption beyond 2~$\mu$m preventing the extension of SC spectra above this limit, and this material aspect is therefore particularly crucial in the mid-infrared regime. Using non-silica glasses with different refractive indices and a \jdnote{nanostructured core design,} we have fabricated a graded-index fiber with extended transmission window \jdnote{ and order of magnitude enhanced nonlinearity compared to silica fibers.} We have then reported periodic spatio-temporal instabilities and the generation of a broadband supercontinuum spanning two-octave up to the mid-infrared. We performed a detailed study of the supercontinuum generation process as a function of the pump wavelength and pump power and our results shows evidence of nonlinear beam cleaning.  \jdnote{Experimental results were compared with spatio-temporal numerical simulations, and the remarkable agreement obtained over the full two-octave bandwidth allows us to confidently interpret the dominant physical mechanisms underlying the supercontinuum broadening, even in the complex multimode regime.  Some limitations to the achievable spectral bandwidth remain, associated with the strong OH attenuation at longer wavelengths.  We believe that improving the purification process of the constituent glasses will allow extension of the SC spectrum even further, and opening up a new avenue for the development of high-brightness supercontinuum sources in the mid-infrared.}

\section*{Acknowledgements}
\noindent ZE acknowledges the Horizon 2020 Framework Programme (722380); LS acknowledges the Faculty of Engineering and Natural Sciences graduate school of Tampere University. JD acknowledges the French Agence Nationale de la Recherche (ANR-15-IDEX- 0003, ANR-17-EURE-0002,ANR-20-CE30-0004). GG acknowledges the Academy of Finland (333949, Flagship PREIN 320165). 

\section*{Author contributions}
ZE performed all the experiments. ZE, AF, DP, MK, and RB designed and fabricated the fiber. LS and ZE performed the numerical simulations. ZE, LS, JD and GG performed the data analysis and interpretation, and all authors participated in the writing of the manuscript. GG planned the research project and provided overall supervision.

\section*{Competing interests}
The authors declare no competing interests.

\section*{Methods}
\subsection*{Fiber fabrication}
The GRIN fiber was fabricated using a stack-and-draw method and nanostructured core approach as described in Ref. \cite{buczynski2015optical}, similar to those used to fabricate photonic crystal fibers. The gradient index profile was obtained by stacking two types of lead-bismuth-gallate glass rods (PBG81 and PBG89) with external diameter of 5~cm and distinct refractive indices. The hexagonal structural arrangement of the rods was optimized using a stochastic simulated annealing optimization inspired from the Maxwell-Garnet model \cite{sihvola1999electromagnetic}. The stacked structure was subsequently drawn in order to fuse the glass rods together to a diameter of about 5 mm. The preform structure was then placed in an external tube acting as the fibre cladding and drawn to a diameter of 125 um. After the drawing process, the sub-wavelength diameter of the rods effectively yields a gradient index in the core area \cite{hudelist2009design,buczynski2015optical}. 

\subsection*{Experimental setup}
The femtosecond laser beam was focused into the fiber using a 5~cm focal length MgF$_2$ plano–convex lens resulting in a beam radius of 25~µm (1/e$^2$ intensity) at the input facet. The fiber holder was placed on a  three-axis precision translation stage to control the input coupling conditions. The fiber was laid straight without any bending or other stress. A dichroic filter was used to select the signal or idler depending on the pump wavelength, and a spectral longpass filter was inserted to attenuate the OPA pump residue. The maximum throughput was about 45\% (calculated as a ratio of output to input power and including coupling efficiency, Fresnel reflection losses, and attenuation along the fiber). Two optical spectrum analyzers (AQ6315B and AQ6376) were used to measure the SC spectrum in the 350--1700~nm and 1500--3400~nm wavelength range, respectively. Spatial intensity measurements were performed in the far field at a distance of 2.5 cm using a beam profiling camera (Pyrocam IIIHR). 

\subsection*{Noise measurements}
 To characterize the SC pulse-to-pulse intensity fluctuations, light from the fiber output was collimated and directed to a monochromator to filter out wavelength bands with a bandwidth of 6~nm. This bandwidth was selected to yield sufficient energy to be detected as well as to minimize wavelength-averaging. A MgF$_2$ plano–convex lens with a 10 cm focal length was used to focus light from the monochromator output to a 15 MHz photodetector (PbSe; PDA10D-EC) connected to a fast 1 GHz real-time oscilloscope (LeCroy WaveRunner 6100 A).  \jdnote{Spectral filtering at the monochromator output removed the second-order diffraction of the SC short wavelength components.} The relative intensity noise (RIN) defined as the standard deviation over the mean value was calculated by integrating the voltage of 4000 consecutive pulses after subtracting the noise background. 
 
\subsection*{Geometric parametric instability sidebands}
The frequencies of the geometric parametric instability sidebands $f_m$ of order $m$ where $m=\pm 1, \pm 2, \pm 3...$ are governed by \cite{longhi2003modulational,agrawal2019invite}
\begin{equation}
    (2\pi f_m)^2 = \frac{2\pi m}{z_p\beta_2} - \frac{2n_2 I \omega_0}{c\beta_2}
    \label{eq:1}
\end{equation} 
where $n_2$ is the nonlinear refractive index, $\omega_0$ is the center angular frequency, I is the pump intensity and $\beta_2$ the group velocity dispersion coefficient at $\omega_0$. The parameter $z_p=\pi \rm R/\sqrt{2\Delta}$ is the self-imaging period with $\Delta=(n_{co}-n_{cl})/n_{co}$ and R the fiber core radius. For our fiber at 1700 nm $n_{co}=1.885$, $n_{cl}=1.866$, $\Delta=0.0101$, R=40 µm, $z_p=0.89$~mm, $\beta_2=5.93 \times 10^{-26}$~s$^2$m$^{-1}$, $n_2=1.95 \times 10^{-19}$~m$^2$W$^{-1}$.

\subsection*{Spatio-temporal numerical simulations}
We simulate the propagation in the fiber using the 3+1D generalized nonlinear Schrödinger equation which is an extension of the Gross–Pitaevskii equation \cite{yu1995spatio,raghavan2000spatiotemporal}
\begin{equation}
\begin{split}
    \partial_z A - i\frac{1}{2\beta_0}\nabla_T^{2}A - i\hat{D}A + i\frac{\beta_0\Delta}{R^2}r^2A = & i\gamma(1+\tau_s\partial_t)[(1-f_R)|A|^2A \\
    &+ f_RA\int_{-\infty}^t h_R(\tau)|A(t-\tau)|^2 \text{d}\tau],
    \label{eq:3dnlse}
\end{split}
\end{equation}
where the field envelope $A$ is expressed in $\sqrt{W}$, $r^2 = x^2 + y^2$, $\nabla_T^{2} = \partial_x^2+\partial_y^2$ is the transverse Laplacian, $\hat{D} = \sum_{n\geq2}(i\partial_t)^n\beta_n/n!$ is the dispersion operator expanded in terms of Taylor-series coefficients, and $\tau_s$ is the shock term. For completeness, the Raman contribution $h_R$ was included and modeled as a delayed response using the conventional form
\begin{equation}
    h_R(\tau) = \left(\tau^2_1 + \tau^2_2\right)\tau_1e^{-\frac{\tau}{\tau_2}}\sin(\tau/\tau_2),
    \label{eq:Raman}
\end{equation}
with $\tau_1$=5.5~fs, $\tau_2$=32~fs, and $f_R$ = 0.05 \cite{sobon2014infrared} but its effect of the propagation dynamics was found to be negligible (see Supplementary information). 

The simulations consider pulses of 350~fs duration (FWHM) and a Gaussian temporal intensity profile. Shot noise was added via one-photon-per-mode with random phase in the frequency domain \cite{dudley2006supercontinuum}. We also included a relative intensity noise (0.2\% of the peak intensity) in the temporal domain corresponding to the intensity fluctuations of the OPA. The refractive index profile is defined along the radial coordinate by $n(r)=n_{co}-ar^2$ for $r\leq \rm R$ and $n(r)=n_{cl}$ for $r>\rm R$, where $n_{co}$ and $n_{cl}$ are the refractive at the center of the core and in the cladding, respectively. $\Delta$ is the relative refractive index difference between the core and the cladding and R is the core radius.

The Taylor-series expansion coefficients for the dispersion operator are calculated from the dispersion profile of the fundamental mode and at 1700~nm they are
$\beta_2 =  5.93 \times 10^{-26}$~s$^2$m$^{-1}$,
$\beta_3 =  3.10 \times 10^{-40}$~s$^3$m$^{-1}$,
$\beta_4 = -7.35 \times 10^{-55}$~s$^4$m$^{-1}$,
$\beta_5 =  2.15 \times 10^{-69}$~s$^5$m$^{-1}$,
$\beta_6 = -3.60 \times 10^{-84}$~s$^6$m$^{-1}$,
$\beta_7 =  2.84 \times 10^{-99}$~s$^7$m$^{-1}$, and
$\beta_8 =  1.35 \times 10^{-115}$~s$^8$m$^{-1}$.
The relative refractive index difference between the core and the cladding $\Delta = 0.0101$, and the nonlinear refractive index $n_2 =  1.92 \times 10^{-19}$~m$^2$W$^{-1}$.

At the center wavelength of 2500~nm, the dispersion coefficients are 
$\beta_2 = -1.15 \times 10^{-25}$~s$^2$m$^{-1}$,
$\beta_3 =  7.35 \times 10^{-40}$~s$^3$m$^{-1}$,
$\beta_4 = -1.74 \times 10^{-54}$~s$^4$m$^{-1}$,
$\beta_5 =  3.60 \times 10^{-69}$~s$^5$m$^{-1}$,
$\beta_6 = -4.60 \times 10^{-84}$~s$^6$m$^{-1}$,
$\beta_7 =  2.78 \times 10^{-99}$~s$^7$m$^{-1}$, and
$\beta_8 =  2.19 \times 10^{-115}$~s$^8$m$^{-1}$.
and the relative refractive index difference between the core and the cladding $\Delta = 0.0096$.

The spatial intensity distribution of the beam is taken to be Gaussian with an input beam radius of 25~$\mu$m (1/e$^2$ intensity radius). In order to mimic imperfections in the input beam and in the free-space coupling to the fiber, we use a similar approach as in Ref.~\cite{krupa2017spatial} and apply to the spatial amplitude distribution a multiplicative phase mask where a random phase shift from (0 to $\pi$) is added to each spatial grid points. During the very first few centimeters of propagation, large angular frequencies leak to the cladding and a supergaussian filter is applied to absorb the field at the boundaries of the spatial window. A supergaussian spectral short-pass filter is also applied to reproduce the strong attenuation of spectral components beyond 2800~nm. The total losses resulting from leakage to the cladding and infrared absorption are 3-5 dB depending on the injected power and pump wavelength. The peak power used in the simulations are adjusted to match the experimentally measured energy at the fiber output.

The simulation grid consists of 16384 spectral/temporal grid points with a temporal window size of 20~ps and 64$\times$64 spatial points with a window size of 160~$\mu$m. A step size of 37~$\mu$m was used in the propagation direction. The beam profiles in the far-field are calculated as the Fourier transform from the near-field at the fiber output. The simulated spectra are averaged over 10 realizations with different noise seeds, and the spectra are convolved with a super Gaussian filter with 2~nm bandwidth which corresponds to our experimental resolution.



\clearpage
\bibliographystyle{MSP}
\bibliography{ref}




\end{document}